# E-NET MODELS OF A SOFTWARE SYSTEM FOR WEB PAGES SECURITY

**Nikolai Todorov Stoianov, Veselin Tsenov Tselkov**

***Abstract:*** *This paper presents solutions for cryptography protection for web pages. The solutions comprise the authors' experience in development and implementation of systems for information security in the Automated Information Systems of Bulgarian Armed Forces. The architecture, the models and the methods are being explained.*

## 1. INTRODUCTION

With the increase of the possibilities of Internet/Intranet services and their daily use in the corporative information systems the problem of the information security is being very topical. One of the most frequently used services is access to web pages. The advantages of using it are:
- easy to use;
- comparatively inexpensive;
- very popular.

Except for these advantages this service also has many disadvantages:
- usually attacked by hackers;
- everyone could print and copy pages;
- lack of security communication;
- there is no strict identification and authentication on web server.

Some of the existing solutions of the web security are:
- using the software system for public key information security;
- MS Internet Explorer, Netscape Navigator, etc.;
- Using SHTML;
- integration of digital certificate in web software (Baltimore, Microsoft, Netscape, Entrust etc.) [1,2,3].

These means give the opportunity to use widespread software such as MS Internet Explorer, MS Word, MS Excel. But it is not possible to:
- control access to all web pages;
- cipher pages;
- keep logs;
- analyze the state of the page;
- generate reports for accomplished actions.

Consequently it is necessary to create a software system for web security which should use all positive characteristics of this service, and to eliminate the disadvantages related with the information security.

## 2. CSSW ARCHITECTURE

CSSW is a solution for cryptography software for information security in Intranet.

CSSW is Windows based program. This software uses secret key algorithms and public key algorithms [4,6]. It gives possibilities for:
- identification, authentication and user control;
- ciphering data in workstation;
- generation and verification of signature;
- log of accomplished actions;
- user's interface and implementing user's applications.

CSSW is an open system for creating applications for data security. Such applications are protection of data on hard disk, directory and file; e-mail security; web pages security; clipboard security; database security. All CSSW applications are based on Microsoft standard and they are easy to integrate in Microsoft products (MS Outlook, MS Internet Explorer, MS Word etc.).

## 3. PURPOSE, FUNCTIONAL POSSIBILITIES AND ARCHITECTURE OF A SOFTWARE SYSTEM FOR WEB PAGES SECURITY

The purpose of a software system for web security is to provide the creation, ciphering and access to security web pages by using the standard software and CSSW functions. When creating, ciphering and accessing crypted pages the system should secure it by forbidding unauthorized user to change, read or remove the pages.

The system should work with MS Internet Explorer and CSSW, which determines its module structure. The architecture of the software system for web security consists of:
- module for end user with Plug-in for MS Internet Explorer and SecWeb application;
- module for creating and ciphering security web pages;
- module for control and monitoring;
- module for distribution and control of cryptography keys.

There are different kinds of systems for accessing and reading web pages. The scheme for access to crypted web pages is shown on fig.1 [5].

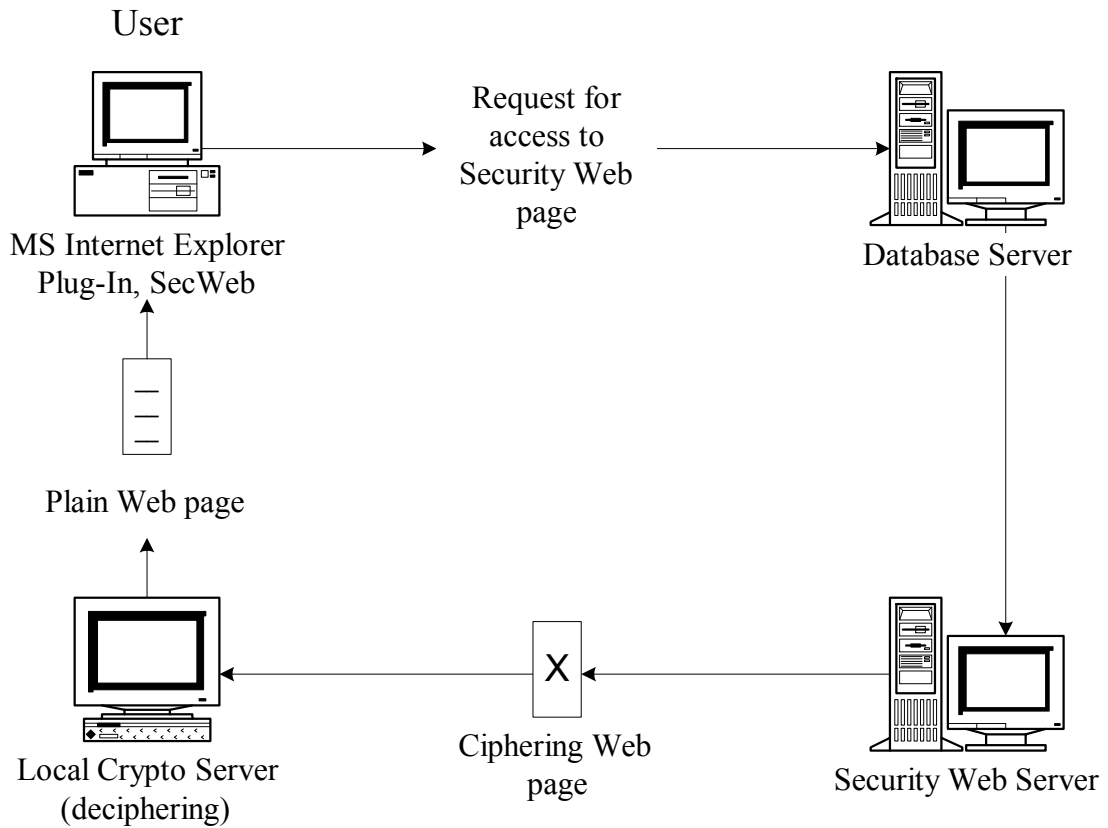

Fig. 1 Scheme for access to security web pages

## 4. E-NET MODEL FOR CRYPTOGRAPHY WEB SECURITY

The functional possibilities for user are:
- request for access;
- receiving an answer;
- analysis of the answer;
- creating plain web pages;
- ciphering web pages;
- publishing web pages;
- deciphering web pages;
- reading web pages;
- quit from the system.

The E-net model ENC=<B, Bp, Br, T, F, H, Mo> for creating and preparing a web page is shown on fig.2.

B={bp1, br1…br5, b1…b5} is the set of model ENC's places.

T={t1…t6} is the set of transitions.

The places and the transitions in the model are in the sense of general places and general transitions.

The relations between places and transitions (functions F and H) are shown on fig.2.

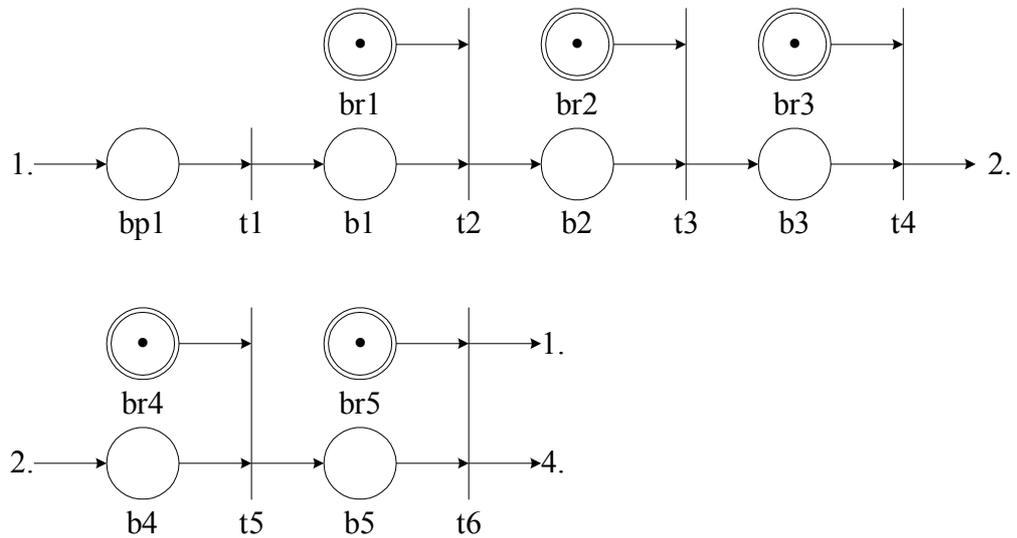

Fig. 2 ENC model for creating and preparing a web page

Places of the model ENC.
Places of the model describe the state and interaction of the users with the system for web security.
Bp={bp1} is the set of peripheral positions and kernel has appeared in bp1 just when a user wants to utilize the system.
Br={br1…br5} is the set of permissive places respectively at transitions t2, t3, t4, t5 and t6.
Transitions of the model ENC.
Transitions of ENC simulate:
- t1 – creating plain web pages;
- t2 – setting access rights for web pages;
- t3 – ciphering web pages;
- t4 – publishing web pages on Security Web Server;
- t5 – generating access list for users;
- t6 - exit from the system.

Kernels of the model ENC.
Kernels' descriptions in different model ENC's places correspond to the input (output) parameters of according transitions.

The E-net model ENA=<B, Bp, Br, T, F, H, Mo> for access to a security web page is shown on fig.3.
B={bp1, br1…br6, b1…b9} is the set of model ENA's places.
T={t1…t9} is the set of transitions.
The places and the transitions in the model are in the sense of general places and general transitions.
The relation between places and transitions (functions F and H) are shown in the fig.3.

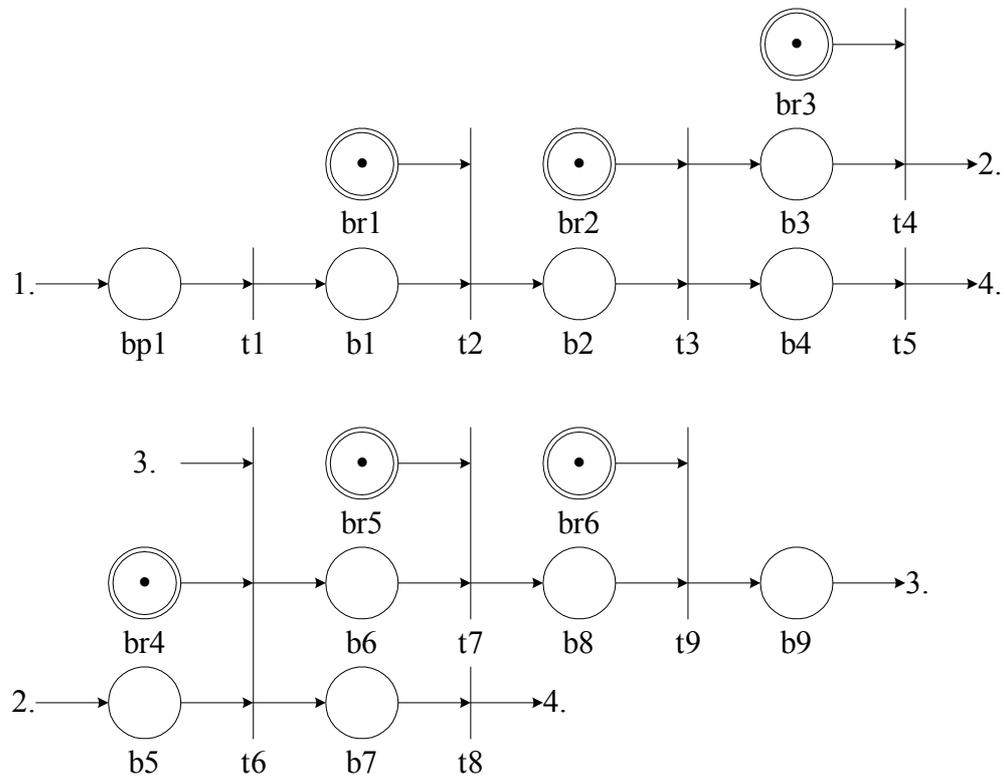

Fig. 3 ENA model for access to a security web page

Places of the model ENA.

Places of the model describe the state and interaction of the users with the system for web security.

Bp={bp1} is the set of peripheral positions and kernel has appeared in bp1 just when a user wants to utilize the system.

Br={br1…br6} is the set of permissive places respectively at transitions t2, t3, t4, t6, t7 and t9.

Transitions of the model ENA.

Transitions of ENA simulate:
- t1 – simulating request for access to standard web browser;
- t2 – requesting for access to security mode;
- t3 – verification of user's rights;
- t4 – replying to request for access to security mode;
- t5, t8 – exit from the system;
- t6 – requesting for access to security web page;
- t7 – replying to request for access to security web page;
- t9 – deciphering security web page and shown in SecWeb.

Kernels of the model ENA.

Kernels' descriptions in different model ENA's places correspond to the input (output) parameters of according transitions.

## 5. SUMMARY
On the bases of the E-net models ENC and ENA algorithms and software system for cryptography Web pages security is being realized. The suggested models have been approbated in Center for Information Security of Defense Advanced Research Institute.